\documentclass[aps,prd,superscriptaddress,amsmath,amsfonts,groupedaddress,10pt,english]{revtex4}
\usepackage{amssymb,amsmath,epsfig}
\usepackage{graphicx}
\usepackage{float}
\usepackage[utf8]{inputenc}
\usepackage[caption=false]{subfig}
\begin{document}
\title{Traversable wormholes with static spherical symmetry and their stability in higher-curvature gravity}

\author{M. Ilyas}
\email{ilyas\_mia@yahoo.com}
\affiliation {Institute of Physics, Gomal University, Dera Ismail Khan, 29220, KP, Pakistan}
\author{Kazuharu Bamba}%
\email{bamba@sss.fukushima-u.ac.jp}
\affiliation{Faculty of Symbiotic Systems Science,
Fukushima University, Fukushima 960-1296, Japan}

\begin{abstract}
The solutions of traversable wormholes and their geometries are investigated in higher-curvature gravity with boundary terms for each case under the presence of anisotropic, isotropic and barotropic fluids in detail. For each case, the effective energy-momentum tensor violates the null energy condition throughout the wormhole throat. The null and weak energy conditions are also analyzed for ordinary matters. The regions that physically viable wormhole solutions can exist are explicitly shown. Furthermore, it is found that the range of the viable regions exhibits an alternating pattern of expansion and contraction. The present analyses can reveal the regions in which traversable wormholes can be constructed for anisotropic, isotropic and barotropic fluids cases with incorporating realistic matter contents, leading to fundamental physics insights into the feasible construction of wormholes in higher-curvature gravity with boundary term. The main achievements of this work, in contrast to previous studies, are its thorough investigation of traversable wormholes within the framework of higher-curvature gravity with boundary terms, its extensive consideration of various fluid types, and the explicit identification of regions where stable wormhole solutions can exist.
\end{abstract}

\maketitle

\section{Introduction}
Over the last decade, significant advancements have occurred in the domains of cosmology and astrophysics, particularly in unraveling the enigma surrounding the universe's accelerated expansion \cite{1,2,3,4}. Empirical findings sourced from various observational avenues such as supernova type Ia \cite{1}, cosmic microwave background \cite{2}, large-scale structure \cite{3}, weak lensing \cite{4}, and other observational methodologies have consistently converged on the presence of a mysterious influence, referred to as dark energy (DE). This concealed force, marked by a substantial negative pressure, stands as the impetus driving the cosmic acceleration. While the simplest interpretation attributes DE to the cosmological constant, this explanation encounters persistent obstacles such as the fine-tuning predicament and the cosmic coincidence paradox.

In response, two distinct propositions have emerged to shed light on the phenomenon of cosmic acceleration. The first approach involves the integration of diverse scalar fields within the framework of Einstein's gravity. These encompass entities like k-essence, quintessence, the cosmological constant \cite{5}, and the Chaplygin gas, among others. These scalar fields strive to offer alternative depictions of DE. The second avenue of inquiry concentrates on the modification of the Einstein-Hilbert action, giving rise to the formulation of modified theories of gravity. Illustrations of these modifications encompass $f(R)$ gravity \cite{7}, $f(R, T)$ gravity \cite{8} (where $R$ signifies the Ricci scalar and $T$ represents the energy-momentum tensor trace), Brans-Dicke theory \cite{9}, Gauss-Bonnet (GB) theory \cite{10}, and more. These adapted theories of gravity endeavor to account for the observed cosmic acceleration by reshaping gravitational dynamics. For comprehensive insights into the subject of dark energy and modified theories of gravity aimed at elucidating late-time cosmic acceleration, refer to comprehensive reviews such as those presented in References \cite{10a1,10a2,10a3,10a4,10a5,10a6,10a7,10a8,10a9,10a10,10a11,10a12,10a13,10a14}.

By investigating scalar field theories and modified gravity theories, researchers aim to unravel the mechanisms underlying the accelerated expansion of the universe. This pursuit not only addresses the challenges associated with the cosmological constant but also enhances our understanding of dark energy \cite{5,7,8,9,10}. The exploration of these theories expands our knowledge of the fundamental forces shaping the cosmos \cite{5,7,8,9,10}. The theory of modified Gauss-Bonnet (GB) gravity, alternatively referred to as higher-order Gauss-Bonnet gravity (where $G$ signifies the GB curvature invariant), is derived by introducing an arbitrary function associated with higher-order Gauss-Bonnet gravity into the Einstein-Hilbert action \cite{11}. For an exploration of modified theories of gravity, consult Ref.~\cite{11a}. This adapted theory effectively encompasses diverse aspects of cosmology, encompassing late-time cosmic acceleration, the transition from deceleration to acceleration, and successfully clears tests within the solar system, contingent on a judicious selection of higher-order Gauss-Bonnet gravity models \cite{12,13,14}. Scholars have formulated cosmologically tenable models within the framework of this gravitational theory \cite{15}, while others have explored realistic variations \cite{16}. Certain investigations have centered on cosmological solutions, particularly within the $\Lambda$CDM model, concluding that both the inflationary epoch and the era of dark energy can be effectively addressed through this theoretical approach. In essence, modified Gauss-Bonnet gravity, or higher-curvature gravity with a boundary term, has been established through the integration of an arbitrary function into the Einstein-Hilbert action. It has demonstrated its capacity to elucidate a multitude of phenomena in cosmology, exhibiting success in solar system tests and furnishing viable models for cosmic acceleration and the universe's evolution \cite{11,12,13,14,15,16}.

A wormhole (WH) proposes a speculative conduit or connection that hypothetically links distinct universes (inter-universe WH) or distant regions within the same universe (intra-universe WH) \cite{18}. When such a wormhole allows for passage in both directions, it is referred to as a traversable wormhole \cite{18}. The concept of traversable wormholes was initially introduced through a two-dimensional embedding diagram depicting the Schwarzschild wormhole \cite{18}. Non-traversable solutions for the Schwarzschild wormhole were explored in existing literature \cite{19}. The potential for traversable wormholes was investigated, revealing the necessity of exotic matter, which violates the null energy condition (NEC), for their feasibility \cite{20}. The quest for realistic models that satisfy energy conditions has presented an intriguing challenge. Various approaches have been explored, including dynamical wormhole solutions \cite{21}, brane wormholes \cite{23}, and generalized Chaplygin gas \cite{24}, all aimed at minimizing the violation of NEC. Modified gravity theories, including GB theory \cite{25}, BD theory \cite{26}, Horava gravity \cite{27}, hybrid metric-Palatini gravity \cite{28}, $f(R)$ gravity \cite{29}, and others, have been examined as potential frameworks supporting wormholes (WHs). WH solutions that adhere to the weak energy condition (WEC) near the throat have been explored within $R^{-1}$ and $R^2$ gravity \cite{30}. WHs in generalized teleparallel gravity have been discussed, confirming the fulfillment of energy conditions at the throat and in its vicinity \cite{31}. Traversable WH solutions in $f(R, T)$ gravity have been scrutinized, revealing NEC compliance for both isotropic and barotropic equations of state (EoS). Physically plausible WH solutions for both barotropic and isotropic scenarios have been identified. However, in generalized teleparallel gravity, the NEC is violated for anisotropic WHs. Specific solutions for $f(R) = R + a R^2$ have been examined, determining the existence of physically acceptable WHs only within specific parameter regions for the barotropic case.

Extensive scrutiny of WH solutions within the framework of noncommutative geometry has been undertaken. The exploration of WH solutions employing noncommutative geometry has uncovered asymptotically flat and nonflat solutions in four and five dimensions, respectively, while higher-dimensional WH solutions remain elusive. Noncommutative WHs utilizing the power law form $f(R) = a R^n$ have been investigated, demonstrating a decrease in throat radius with increasing power of $R$. WH solutions within the realm of generalized teleparallel gravity have also been investigated within the same noncommutative backdrop \cite{37}. Within the context of modified theories, much focus has been placed on investigating wormhole solutions with zero tidal force, simplifying field equations and avoiding complications. However, this study delves into static wormhole solutions within higher-order Gauss-Bonnet gravity, incorporating non-zero tidal forces. Specifically, anisotropic, isotropic, and barotropic fluids are considered, adopting a specific viable model. The solutions are examined by assessing the null energy condition (NEC) and the weak energy condition (WEC) in each scenario. In recent studies exploring the intriguing realm of wormholes, researchers have investigated the geometry, stability, and echoes of galactic wormholes \cite{37a} as well as echoes emanating from braneworld wormholes \cite{37b}. In our investigation of wormhole existence and their implications within extended gravity theories, we rely on key studies such as those examining the General relativistic Poynting-Robertson effect in static and spherically symmetric cases \cite{37c, 37d}, reconstructing wormhole solutions in curvature-based Extended Theories of Gravity \cite{37e}, and analyzing epicyclic frequencies in various wormhole geometries \cite{37f, 37g}. These references provide crucial insights and analytical tools for our research.

The structure of the paper is organized as follows: In Section II, we discuss the wormhole in General Relativity (GR). In Section III, we delve into the discussion of higher-order Gauss-Bonnet gravity, including its action, field equations, and matter content. Moving on to Section IV, we explore the intricacies of wormhole geometry along with anisotropic matter content. In Section V, we analyze the matter contents of perfect fluids in the context of wormhole geometry. Section VI is dedicated to the equation of state for barotropic fluid scenarios. Shifting our focus to Section VII, we perform a comprehensive stability analysis of the solutions we've presented. Finally, in Section VIII, we conduct a comparative analysis of wormholes in GR and higher-order Gauss-Bonnet gravity. In the concluding section, we consolidate our results and discussions, weaving together insights and future directions into a cohesive tapestry that enhances our understanding of modified gravity, wormholes, and the intriguing interplay of energy conditions.

\section{Wormholes in General Relativity with Anisotropic Phantom Matter}
In this section, we explore the possibility of wormholes within the framework of GR when considering anisotropic phantom matter. This investigation serves as a foundational exploration before we delve into wormholes in higher-order Gauss-Bonnet gravity. We aim to analyze the feasibility of traversable wormholes under the influence of phantom-like exotic matter.
\subsection{The Wormhole Metric}
The interior wormhole spacetime is described by the following metric:
\begin{equation}
d{s^2} = {e^{a(r)}}d{t^2} - {e^{b(r)}}d{r^2} - {r^2}\left( {d{\theta ^2} + {{\sin }^2}\theta d{\phi ^2}} \right).
\label{metricwormhole}
\end{equation}
Here, $a(r)$ representing the redshift function and $b(r)=(1 - \frac{\beta(r)}{r})^{-1}$ in which $\beta(r)$ representing the shape function. The wormhole's throat is situated at $b(r_0) = r_0$. To ensure traversability, we must guarantee the absence of horizons, defined as the surfaces where $e^{a}$ approaches zero. Therefore, $a(r)$ must remain finite everywhere. Additionally, we impose the condition $1 - \frac{b}{r} > 0$.
\subsection{Stress-Energy Tensor for Phantom Matter}
The stress-energy tensor components for the anisotropic phantom matter within the wormhole are given by:
\begin{eqnarray}
\rho(r) &=& \frac{b'}{r^2} \label{rhoWH}, \\
p_r(r) &=& \left[-\frac{b}{r^3} + 2\left(1 - \frac{b}{r}\right)\frac{a'}{r}\right] \label{prWH}, \\
p_t(r) &=& \left(1 - \frac{b}{r}\right)\Bigg[a'' + (a')^2 - \frac{b'r - b}{2r(r - b)}a' - \frac{b'r - b}{2r^2(r - b)} + \frac{a'}{r}\Bigg] \label{ptWH}.
\end{eqnarray}
The conservation of the stress-energy tensor, $T^{\mu\nu}{}_{;\nu} = 0$, yields the following relationship:
\begin{equation}
p_r' = \frac{2}{r}(p_t - p_r) - (\rho + p_r)a' \label{prderivative}.
\end{equation}
\subsection{Violation of Null Energy Condition (NEC)}
Traversable wormholes and phantom energy are closely linked through the violation of the null energy condition (NEC), defined as $T_{\mu\nu}k^{\mu}k^{\nu} \geq 0$, where $k^\mu$ is any null vector and $T_{\mu\nu}$ is the stress-energy tensor. For phantom energy, characterized by an equation of state $\omega = p/\rho$ with $\omega < -1$, the NEC is indeed violated, i.e., $p + \rho < 0$.
For our wormhole spacetimes, we consider an orthonormal reference frame with $k^{\hat{\mu}} = (1, \pm 1, 0, 0)$, leading to the following NEC evaluation at the throat:
\begin{equation}\label{NECthroat}
T_{\hat{\mu}\hat{\nu}}k^{\hat{\mu}}k^{\hat{\nu}} = \left[\frac{b'r - b}{r^3} + 2\left(1 - \frac{b}{r}\right)\frac{a'}{r}\right].
\end{equation}
By considering the flaring out condition of the throat, $(b - b'r)/2b^2 > 0$, and accounting for the finite character of $a(r)$, we verify that the NEC is indeed violated when evaluated at the throat, i.e., $T_{\hat{\mu}\hat{\nu}}k^{\hat{\mu}}k^{\hat{\nu}} < 0$. Matter that violates the NEC is referred to as {\it exotic matter}.\\

\subsection{Anisotropic Case}

In the study of wormholes, it is essential to consider scenarios where the pressure distribution is anisotropic, meaning that the radial pressure \(P_r\) and the tangential pressure \(P_t\) are not equal. The anisotropic case leads to the following differential equation, which relates the energy density \(\rho\) and \(P_r\):

\begin{equation}
\rho + P_r = \frac{r\left(ra' + b'\right) - b\left(ra' + 1\right)}{r^3}
\end{equation}

This equation characterizes the interplay between the energy content and the pressure distribution within the wormhole. To gain insights into the physical consequences of this scenario, we use Eq. (\ref{eq:beta}) and Eq. (\ref{1b}) to investigate the behavior of the null energy condition (NEC).

The NEC is a fundamental principle in general relativity that requires the energy density at any point in spacetime to be non-negative. Violation of the NEC signifies exotic matter distribution, which is often associated with theoretical constructs like wormholes.
\begin{figure}[H]
  \centering
    \includegraphics[width=0.5\linewidth]{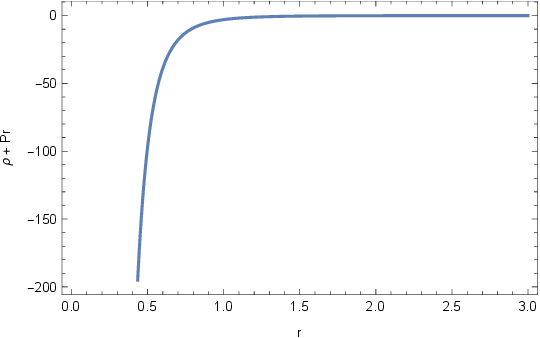}
  \caption{Anisotropic case}
  \label{gr1}
\end{figure}

Figure \ref{gr1} illustrates the behavior of the NEC in the anisotropic case. Upon close examination of the plot, we observe that the NEC is indeed violated in this scenario. This violation implies the presence of exotic matter, which challenges the conventional energy conditions required for stable and traversable wormholes within the framework of general relativity.

The violation of the NEC in the anisotropic case raises intriguing questions about the nature of exotic matter and its role in the stability and traversability of wormholes. Further exploration of alternative matter distributions and possible modifications to the theory of gravity are warranted to understand the physical implications of such violations and their potential consequences for our understanding of spacetime structures.

\subsection{Isotropic Case}
\begin{figure}[H]
  \centering
    \includegraphics[width=0.5\linewidth]{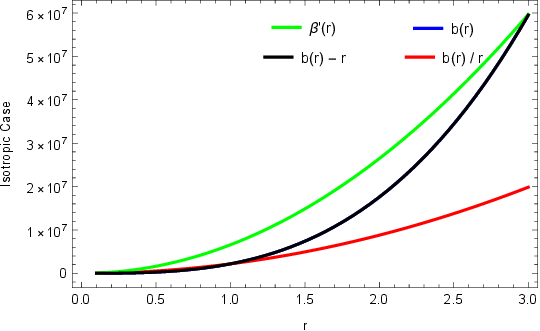}
  \caption{Isotropic case}
  \label{gr2}
\end{figure}

The isotropic case in the context of wormholes involves the following differential equation:
\begin{equation}
P_r - P_t = \frac{r\left(2\left(b' - r^2a''\right) + ra'\left(b' + 2\right) - r^2a'^2\right) + b\left(2r^2a'' + r^2a'^2 - 3ra' - 6\right)}{4r^3} = 0
\end{equation}
We aim to solve for the shape function \(b\) under this isotropic pressure distribution, where \(b\) characterizes the radial dependence of the wormhole throat's radius. However, upon solving for \(b\) as shown in above Equation, we plot the obtained solution in Fig. \ref{gr2}.
\begin{equation}
b = \frac{r}{\epsilon^2\left(2r + \epsilon\right)^6}\left[r^8\left(14208 + c_1\epsilon^2e^{\frac{\epsilon}{r}}\right) + 14208r^7\epsilon + 7104r^6\epsilon^2 + 2432r^5\epsilon^3 + 720r^4\epsilon^4 + 224r^3\epsilon^5 + 64r^2\epsilon^6 + 12r\epsilon^7 + \epsilon^8\right]
\end{equation}
The graph in Fig. \ref{gr2} clearly illustrates the behavior of the shape function in the isotropic case. Upon close examination of this plot, it becomes evident that the wormhole conditions, specifically the crucial flaring-out condition, are not fulfilled. The flaring-out condition is vital for the stability and traversability of a wormhole, as it ensures that the throat of the wormhole does not collapse but instead widens as we move away from its center.
This result implies that the wormholes described by the considered perfect fluid model, with isotropic pressure distribution, do not meet the fundamental requirement for traversability according to general relativity. These findings are of significant importance in understanding the physical viability of such wormholes and may motivate further investigations into alternative exotic matter distributions or modifications to the theory of gravity to explore the possibility of traversable wormholes in the universe.
\subsection{EoS Case}
\begin{figure}[H]
  \centering
  \includegraphics[width=0.5\linewidth]{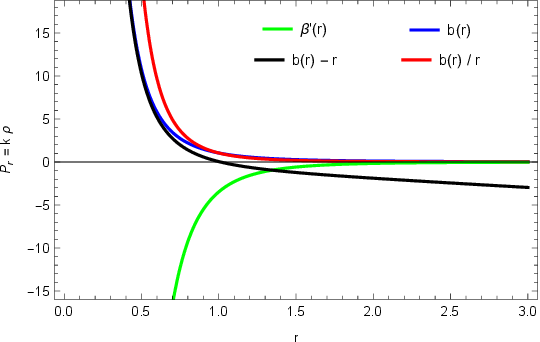}
  \caption{EoS Case: Behavior of the Shape Function \(b(r)\)}
  \label{gr3}
\end{figure}

In the context of wormholes, another important scenario to consider is when the equation of state (EoS) relates the radial pressure \(P_r\) to the energy density \(\rho\) through the parameter \(k\). The EoS case gives rise to the following differential equation:
\begin{equation}
P_r - k\rho = -\frac{rba' - r^2 - a' + krb' + b}{r^3} = 0
\end{equation}
Here, the parameter \(k\) characterizes the specific EoS, which can have significant implications for the wormhole's physical properties. To investigate the radial dependence of the wormhole throat's radius, we aim to solve for the shape function \(b\) by using Eq. (\ref{1b}). The solution for \(b\) in the EoS case is given by:
\begin{equation}
b = \frac{1}{k}r^{-1/k}e^{\frac{\epsilon}{kr}}\left(\epsilon r^{1/k}\left(\frac{\epsilon}{kr}\right)^{1/k}\Gamma\left(-\frac{1}{k}, \frac{\epsilon}{kr}\right) + c_1k\right)
\end{equation}
Here, \(\Gamma\) represents the gamma function.
To gain insight into the physical implications of this solution, we present a graphical representation in Fig. \ref{gr3}, which illustrates the behavior of the shape function \(b\) for the EoS case.
Upon close examination of Fig. \ref{gr3}, it is apparent that the wormhole conditions are not fulfilled in this scenario as well. The flaring-out condition, crucial for the stability and traversability of a wormhole, is not met. This result highlights the significance of the specific equation of state (\(k\)) in determining the wormhole's properties and further underscores the challenges associated with finding traversable wormholes within the framework of this EoS. These findings may stimulate further investigations into alternative matter distributions and gravitational theories to explore the possibility of realizing stable and traversable wormholes.\\

It is worth noting that in the anisotropic case, wormhole solutions are not feasible without the presence of exotic matter. Exotic matter, characterized by its negative energy density and violation of the null energy condition, remains an indispensable ingredient for the stability and existence of wormholes in General Relativity, whether the matter distribution is isotropic or anisotropic.
Furthermore, in the case of a perfect matter fluid with isotropic pressure (e.g., \(P_r = P_t = P\)) and an equation of state where both \(P_r = k\rho\) and \(P_t = k\rho\), it is evident that wormhole solutions cannot be realized with ordinary matter. The exotic matter with properties described in this section is essential for the feasibility of traversable wormholes. This exotic matter violates the null energy condition (NEC) and plays a crucial role in maintaining the wormhole structure. Without such exotic matter, the conditions necessary for a stable and traversable wormhole, as detailed in this section, would not be met. Therefore, the presence of exotic matter, characterized by a specific equation of state and stress-energy tensor components, is a fundamental requirement for the existence of wormholes within the framework of General Relativity.

\section{Higher-order Gauss-Bonnet gravity and field equation}
This section is dedicated to presenting the expanded form of GB gravity, accompanied by its equations of motion and energy conditions.

\subsection{Field equation}
For higher-order Gauss-Bonnet gravity, the usual Einstein-Hilbert action is modified as follows
\begin{equation}\label{action}
S_{GB} = \int {{d^4}x \left( {\frac{R}{2} + f(G)} \right)\sqrt { - g} + {S_M}\left( {{g^{\mu \nu }},\psi } \right)},
\end{equation}
In this equation, ${\kappa ^2} = 8\pi G \equiv 1$, and the symbols $R$, $f$, and ${S_M}({{g^{\mu \nu }},\psi})$ represent the Ricci scalar, an arbitrary function related to the Gauss-Bonnet invariant, and the matter action, respectively. The GB invariant quantity can be expressed as
\begin{equation}
G = R^2 - 4{R_{\mu \nu }}{R^{\mu \nu }} + {R_{\mu \nu \alpha \beta }}{R^{\mu \nu \alpha \beta }},
\end{equation}
where $R_{\mu \nu}$ represents the Ricci tensor and ${R_{\mu \nu \alpha \beta }}$ stands for the Riemannian tensor. By performing variations on the aforementioned action with respect to $g_{\mu \nu }$, the modified field equations for higher-order Gauss-Bonnet gravity can be derived as follows:
\begin{equation}\label{field equation}
{R_{\mu \nu }} - \frac{1}{2}R{g_{\mu \nu }} = T_{\mu \nu
}^{\textrm{eff}},
\end{equation}
where $T_{\mu \nu }^{\textrm{eff}}$ is referred to as the effective energy-momentum tensor and its expression is as follows:
\begin{align}\nonumber
T_{\mu \nu }^{\textrm{eff}}&= {\kappa ^2}{T_{\mu \nu }} - 8\left[
{{R_{\mu \rho \nu \sigma }}} \right. + {R_{\mu \nu }}{g_{\mu \nu }}
- {R_{\mu \nu }}{g_{\mu \nu }} - {R_{\mu \nu }}{g_{\mu \nu
}}+{R_{\mu \nu }}{g_{\mu \nu }}\\\label{field} &+
\frac{1}{2}({g_{\mu \nu }}{g_{\mu \nu }}
 - {g_{\mu \nu }}\left. {{g_{\mu \nu }})} \right]{\nabla ^\rho }{\nabla ^\sigma }f_G
 + \left( {G f_G - f} \right){g_{\mu \nu }},
\end{align}
where the subscript $G$ signifies the differentiation of the respective term with respect to the GB term. Meanwhile, ${T_{\mu \nu }}$ denotes the conventional stress-energy-momentum tensor. Recent advancements in the field of modified gravitational theories have led to the proposal of ghost-free formulations for $F(G)$ gravity and related theories, as discussed in Ref \cite{aaa} and its associated references. Ghosts, which represent unstable modes characterized by negative kinetic terms, have been a subject of concern in certain modified gravity theories. Ghost-free versions have emerged as a potential solution to this issue. Recently, a wormhole solution in Einstein's gravity with two scalar fields that is free of ghosts was proposed in Ref.\cite{aaa1}.

\subsection{Anisotropic Matter Content and Wormhole Geometry}
In this study, our objective is to investigate the impact of anisotropic pressure on the presence of wormhole geometries. To accomplish this, we consider an anisotropic matter distribution as the relativistic source, which can be mathematically described as follows:
\begin{equation}
T^{\mu\nu} = (\rho + P_t) V^\mu V^\nu - P_t g^{\mu\nu} + \Pi X^\mu X^\nu
\end{equation}
In this context, $\rho$, $P_t$, $P_r$, and $P_i$ denote the energy density, tangential pressure, radial pressure, and anisotropic stress component, respectively, of the fluid. Additionally, $V_\mu$ represents the four-velocity of the fluid, while $X_\mu$ corresponds to the radial unit four-vector. These quantities satisfy the conditions $V_\mu V^{\mu} = 1$ and $X_\mu X^{\mu} = 1$ in a co-moving coordinate system.
Furthermore, our focus lies on static spherically symmetric wormholes. In the case of wormhole geometry, we initiate from the line element. The general form of the line element for static spherically symmetric geometry takes the shape of
\begin{equation}
d{s^2} = {e^{a(r)}}d{t^2} - {e^{b(r)}}d{r^2} - {r^2}\left( {d{\theta ^2} + {{\sin }^2}\theta d{\phi ^2}} \right),
\end{equation}
where $a(r)$ and $b(r)$ are arbitrary functions. In the context of the $f(G)$ modified gravity framework, our examination unveils profound alterations in the distribution of energy density ($\rho$), radial pressure ($P_r$), and tangential pressure ($P_t$) in contrast to the expectations of GR. Within the $f(G)$ framework, these physical quantities are influenced not only by the matter content but also by the specific characteristics of the $f(G)$ function and its derivatives. Notably, deviations from GR arise due to the presence of the Gauss-Bonnet term and the additional terms involving $f_G$ (the derivative of $f(G)$ concerning $G$) in our field equations. These deviations reflect the combined impact of matter distribution and the $f(G)$ function on the energy-momentum distribution within the spacetime.
By solving the field equations ${R_{\alpha \beta }} - \frac{1}{2}R{g_{\alpha \beta }} = \kappa T_{\alpha \beta }^{eff}$ where $T_{\mu\nu}^{eff} = T_{\mu\nu}  + {T_{\mu\nu}^{GB}}$ while $\rho^{eff} = \rho  + {\rho _{f(G)}}$, $P_r^{eff} = {P_r} + {P_r}_{f(G)}$ and $P_t^{eff} = {P_t} + {P_t}_{f(G)}$ where
\begin{align}\nonumber
{\rho _{f(G)}} &= \frac{1}{{2{r^5}}}\left[ {8{r^2}b'G'{f^{\prime \prime }}} \right. - 12rbb'G'{f^{\prime \prime }} + 8{r^2}b{f^{\left( 3 \right)}}{{G'}^2} - 8r{b^2}{f^{\left( 3 \right)}}{{G'}^2} + 8{r^2}b{G^{\prime \prime }}{f^{\prime \prime }}\\
& - 8r{b^2}{G^{\prime \prime }}{f^{\prime \prime }} - 8rbG'{f^{\prime \prime }} + 12{b^2}G'{f^{\prime \prime }} + \left. {{r^5}Gf' - {r^5}f} \right],\\
{P_r}_{f(G)}& = \frac{1}{2}\left[ {f - Gf' + \frac{{8\left( {2{r^2} - 5rb + 3{b^2}} \right)a'G'{f^{\prime \prime }}}}{{{r^4}}}} \right],\\\nonumber
{P_t}_{f(G)}& = \frac{1}{2}\left[ {\frac{1}{{{r^4}}}} \right.4\left( {r - b} \right)(r(2r{a^{\prime \prime }}G'{f^{\prime \prime }} + a'( - 3b'G'{f^{\prime \prime }} + 2r{f^{\left( 3 \right)}}{{G'}^2} + 2r{G^{\prime \prime }}{f^{\prime \prime }}) + 2r{{a'}^2}G'{f^{\prime \prime }})\\
& - b(2r{a^{\prime \prime }}G'{f^{\prime \prime }} + 2r{{a'}^2}G'{f^{\prime \prime }} + a'(2r{f^{\left( 3 \right)}}{{G'}^2} + 2r{G^{\prime \prime }}{f^{\prime \prime }} - 3G'{f^{\prime \prime }}))) - Gf'\left. { + f} \right].
\end{align}
Now, let us consider the scenario in the context of GR. In GR, energy density ($\rho$), radial pressure ($P_r$), and tangential pressure ($P_t$) are predominantly determined by the matter content and the metric components of the spacetime, adhering to well-established principles. Under the standard GR framework, the influence of the $f(G)$ function and its derivatives, which were significant in the $f(G)$ framework, are no longer present. Consequently, the energy density, radial pressure, and tangential pressure in GR follow a more conventional pattern, driven solely by the matter content and spacetime geometry. This fundamental distinction emphasizes the unique consequences that modified gravity theories like $f(G)$ can have on the dynamics of gravity, with the precise form of the $f(G)$ function and its derivatives playing a pivotal role in shaping these distinctions from GR.
Subsequently rearranging the results, we derive the explicit expressions for $\rho$, $P_r$, and $P_t$.
\begin{align}\label{fequation}
\rho  &= \frac{f}{2} - \frac{1}{2}G{f_G} + \frac{{b'}}{{{r^2}}} + \frac{1}{{2{r^5}}}\left( {8rb} \right. - 12{b^2}
 - 8{r^2}b' + 12r\left. {bb'} \right){f_G}^\prime  + \frac{1}{{2{r^5}}}\left( { - 8{r^2}b + 8r{b^2}} \right){f_G}^{\prime \prime },\\
P_r &=  - \frac{f}{2} + \frac{1}{2}G{f_G} + \frac{2}{{{r^4}}}\left( {r - b} \right)\left( { - 4r + 6b} \right){f_G}^\prime a'
 - \frac{b}{{{r^3}}} + \frac{{2a'}}{r} - \frac{{2ba'}}{{{r^2}}},\\\nonumber
P_t &=  - \frac{f}{2} + \frac{1}{2}G{f_G} + \frac{1}{{2{r^4}}}( - 8r{\left( {r - b} \right)^2}a'){f_G}^{\prime \prime }\nonumber
 + \frac{1}{{2{r^4}}}(r\left( {1 + ra'} \right)(b - rb' + 2r\left( {r - b} \right)a')\\
& + 2{r^3}\left( {r - b} \right){a^{\prime \prime }}) + \frac{1}{{2{r^4}}}f{G^\prime }( - 4\left( {r - b} \right)a'
(3\left( {b - rb'} \right) + 2r\left( {r - b} \right)a') - 8r{\left( {r - b} \right)^2}{a^{\prime \prime }}),
\end{align}
where
\begin{align}
G&= \frac{1}{{{r^5}}}\left[ { - 8{r^2}b'} \right.a' + 4rb((2 + 3b')a' - 2r{{a'}^2} - 2r{a^{\prime \prime }}) + 4{b^2}( - 3a' + 2r{{a'}^2} + \left. {2r{a^{\prime \prime }})} \right]
\end{align}

In modified theories of gravity, the violation of energy condition (NEC) imposes the condition $T_{\mu \nu }^{eff}{k^\mu }{k^\nu } < 0$,
which is ${\rho ^{eff}} + {p^{eff}} < 0$.
By solving for ${\rho ^{eff}} + {p^{eff}}$, we have
\begin{equation}
\begin{gathered}
  {\rho ^{eff}} + p_r^{eff} = \left[ {\frac{{b'}}{{{r^2}}} - \frac{b}{{{r^3}}} + \frac{{2a'}}{r}\left( {1 - \frac{b}{r}} \right)} \right]
\end{gathered}
\end{equation}
Using the explicit value for $\rho$ and $p_r$ and then solving, we obtain
\begin{equation}
\begin{gathered}
  {\rho ^{eff}} + p_r^{eff} = \frac{1}{{r}^3}\left( {r\beta ' - \beta } \right)
\end{gathered}
\end{equation}
This result is similar to that of $f(R)$ modified gravity. Now if we use the faring out conditions then the above equation becomes
${\rho ^{eff}} + {p^{eff}} < 0$.

For wormhole, ${e^{ - b(r)}} = \left( {1 - \frac{{\beta (r)}}{r}} \right)$. The term $a(r)$ describe the redshift while $\beta (r)$ represent the shape function.

\section{Modified higher-order Gauss-Bonnet gravity Model and Wormhole solution}
We consider the higher-order Gauss-Bonnet gravity model \cite{38}
\begin{equation}\label{model}
f(G) = \frac{{{a_1}{G^{{a_2}}} + {a_3}}}{{{b_1}{G^{{b_2}}} + {b_3}}}
\end{equation}
In this model, the function higher-order Gauss-Bonnet gravity represents the modified gravitational action, and the parameters $a_1, a_2, a_3, b_1, b_2$, and $b_3$ are model parameters.
For higher order effects, we use the above model. The specific form of the higher-order Gauss-Bonnet gravity model presented in Equation (\ref{model}) is aptly suited to fundamental physics due to several key reasons. Firstly, its balanced structure, incorporating terms involving $G^{a_2}$ and $G^{b_2}$ in the numerator and denominator, respectively, ensures theoretical consistency. Moreover, the parameterization provided by $a_1$, $a_2$, $a_3$, $b_1$, $b_2$, and $b_3$ allows tailored adjustments to align with observed phenomena while preserving theoretical integrity. Notably, the model seamlessly transitions to General Relativity (GR) when $a_1$ and $a_3$ are zero, thus upholding GR's predictions in relevant contexts. Additionally, the incorporation of physically motivated components and conservation of energy-momentum reinforces its compatibility with fundamental principles. Lastly, the model's power-law structure makes it well-suited for addressing higher-order effects, making it a robust framework for describing gravitational interactions beyond GR's conventional bounds.

A fundamental property of a wormhole solution impose the condition over shape functions that are the flaring out condition of the throat, mathematically given by $(b - b'r)/b^2 > 0$, and second condition is the throat $r_0$ where $b(r0) = r0$, and the last one condition is $b'(r0) < 1$, which can be imposed for wormhole solutions.

We express the shape function as follows:
\begin{equation}\label{eq:beta}
\beta(r) = r_0 \left(\frac{r_0}{r}\right)^{n},
\end{equation}
where $n$ represents an arbitrary constant. This choice fulfills all the necessary conditions for the existence of a traversable wormhole. The shape function satisfies the condition of being asymptotically flat. At the throat, the condition $\beta(r_0) = r_0$ is inherently satisfied, and for $n > -1$, the condition $\beta'(r_0) < 1$ holds. Various prior studies in the literature have examined wormhole solutions for specific values of $n$ in different theories of gravity.

Further, We assume the red-shift function to be finite,
\begin{equation}\label{1b}
a(r)=-\frac{\epsilon}{r}, \quad\epsilon>0,
\end{equation}
The chosen form of the red-shift function satisfies both the no-horizon condition and exhibits asymptotic flatness at large distances, where $a(r)$ tends towards zero as $r$ approaches infinity.

\begin{figure}[h]
  \centering

  \subfloat[Behavior of the NEC versus $r$ for different values of $n$.]{%
    \includegraphics[width=0.45\linewidth]{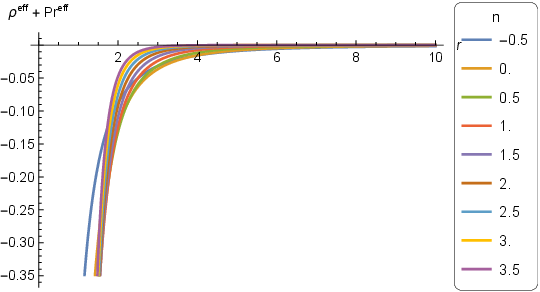}
    \label{fig1}
  }
  \hfill
  \subfloat[Relationship between $\rho$ as functions of $r$ for various values of $n$.]{%
    \includegraphics[width=0.45\linewidth]{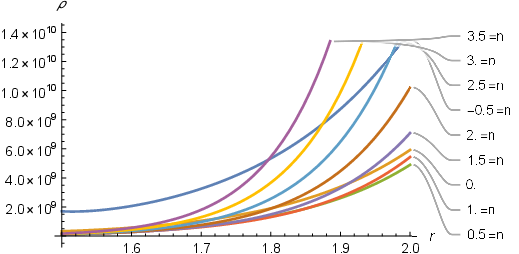}
    \label{fig2}
  }

  \vspace{0.5cm}

  \subfloat[Variation of $\rho+P_r$ with respect to $r$ for different values of $n$.]{%
    \includegraphics[width=0.45\linewidth]{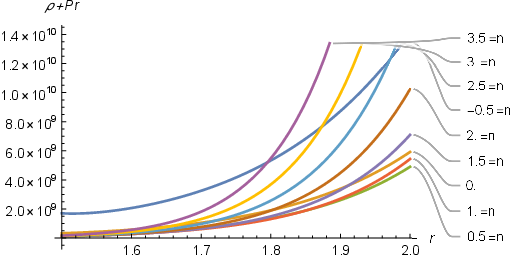}
    \label{fig3}
  }
  \hfill
  \subfloat[Behavior of $\rho+P_t$ versus $r$ for different $n$.]{%
    \includegraphics[width=0.45\linewidth]{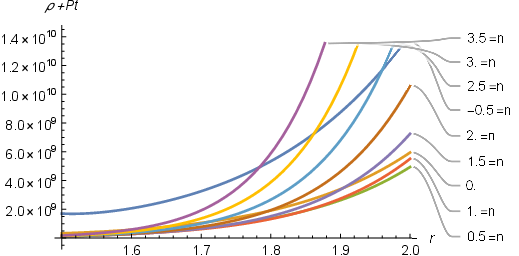}
    \label{fig4}
  }

  \caption{Anisotropic case}
  \label{fig_combined}
\end{figure}

Figure \ref{fig_combined} depict the graphical representations of various energy conditions as functions of the radial coordinate $r$ within the context of anisotropic wormhole solutions in higher-order Gauss-Bonnet gravity. These figures offer valuable insights into the behavior and characteristics of energy within the specific framework of wormholes in this modified theory of gravity.

To generate these, we employ the solutions to the field equations governing anisotropic wormholes in higher-order Gauss-Bonnet gravity. These field equations account for the modifications introduced by higher-order Gauss-Bonnet gravity and provide a comprehensive description of the wormhole geometry. By utilizing these solutions, we can explore the distribution of energy within the wormhole structure.

In addition to the field equations, we incorporate equations (\ref{eq:beta}) and (\ref{1b}) in our analysis. These equations allow us to incorporate the effects of model parameters which play crucial roles in determining the anisotropy and energy conditions of the wormhole. By varying the values of these parameters, we can investigate different scenarios and observe their impact on the energy conditions.

The anisotropic nature of the wormhole solutions in higher-order Gauss-Bonnet gravity introduces interesting features in the energy conditions. These conditions, including the null energy condition (NEC), weak energy condition (WEC), strong energy condition (SEC), and dominant energy condition (DEC), are essential for understanding the physical properties of the wormhole. The graphical representations in Fig. \ref{fig_combined} provide a visual depiction of how these energy conditions evolve as we move along the radial coordinate.

By studying these, we gain valuable insights into the interplay between anisotropy, modified gravity, and energy conditions in the context of wormholes. These investigations contribute to our understanding of the viability and behavior of wormhole solutions within the framework of higher-order Gauss-Bonnet gravity and provide avenues for further exploration in this exciting area of theoretical physics.\\

Figure \ref{fig1} showcases the violation of an energy bound for different values of $n$, where we have chosen $\epsilon=0.01$ and $r_0=1$ as parameters. This violation provides evidence for the existence of physically acceptable wormhole solutions characterized by an anisotropic matter distribution. The careful selection of parameter values allows us to investigate the behavior of the energy conditions specifically for ordinary matter.

Moving on to Fig. \ref{fig2}, \ref{fig3}, and \ref{fig4}, we demonstrate that both the null energy condition (NEC) and weak energy condition (WEC) remain valid throughout the entire evolution for all the considered values of $n$. This observation signifies that the energy conditions, crucial for the physical viability of wormhole solutions, are upheld within the parameter range examined.

This analysis uncovers the presence of viable wormhole solutions within the range $1.5\leq r$, across various values of $n$. Remarkably, as the value of $n$ increases, the region supporting physically acceptable wormhole geometries expands. This expansion suggests that the wormhole structures can persist with the support of ordinary matter as $n$ grows.

These findings highlight the significance of anisotropic matter distributions in enabling the existence of stable wormholes that satisfy the energy conditions. The systematic exploration of parameter values and the study of energy conditions contribute to our understanding of the physical nature of wormholes and aid in identifying feasible scenarios in which such exotic structures can exist within the framework of gravitational theories.
\section{Matter contents of perfect fluids}
\begin{figure}[h]
\centering
\includegraphics[width=0.5\linewidth]{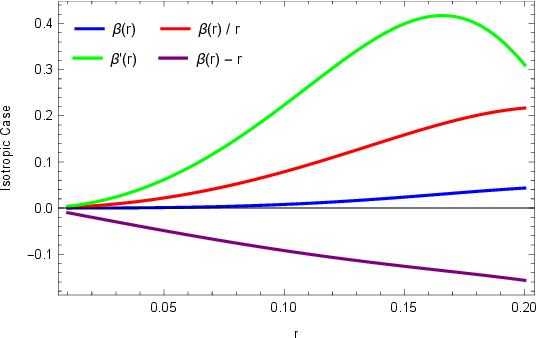}
\caption{Behavior of $\beta(r)$, $\beta(r)-r$, $\frac{\beta(r)}{r}$, and $\beta'(r)$ versus $r$ for isotropic matter content.}
\label{combine fig2}
\end{figure}

In this section, we explore the solution for the perfect matter content, considering the conditions $P_r = P_t$ or $P_r - P_t = 0$. These conditions lead to a third-order non-linear differential equation. Due to the nonlinearity of this equation, it is not possible to derive an analytical solution. Therefore, we employ numerical techniques to solve it, and the obtained results are presented in Fig. \ref{combine fig2}.\\

Figure \ref{combine fig2} illustrates the graphical behavior of $\beta(r)$, which demonstrates that $\left(1-\frac{\beta}{r}\right)$ remains positive throughout the evolution. In we observe that the approximate location of the wormhole throat is around $r_0\approx0.01$, where $\beta(r)-r$ approaches zero. The behavior indicates that the asymptotic flatness condition is satisfied. Furthermore, confirms that $\beta'(r_0)<1$ is upheld.

These findings support our conclusion that the numerical techniques employed successfully provide insights into the behavior of the solution. The graphical representations help us visualize important aspects, such as the positivity of $\left(1-\frac{\beta}{r}\right)$, the location of the wormhole throat, the asymptotic flatness, and the constraint on the derivative of $\beta(r)$.\\

\begin{figure}[h]

  \subfloat[Effective Null Energy Conditions for isotropic matter content.]{\includegraphics[width=0.45\textwidth]{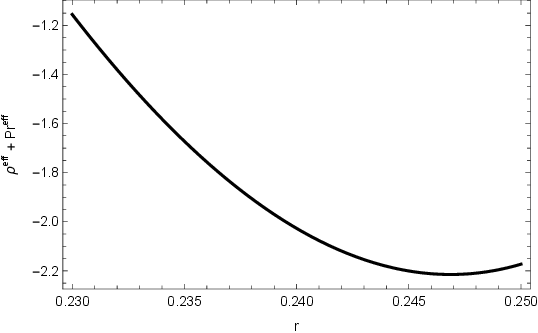}\label{fig6}}
  \hfill
  \subfloat[Behavior of the energy conditions.]{\includegraphics[width=0.45\textwidth]{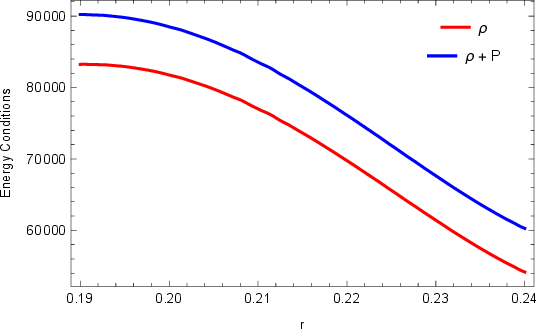}\label{fig:isoropp}}

  \caption{Behavior of energy conditions for isotropic matter content.}
  \label{combine fig3}
\end{figure}

The energy conditions for the case of isotropic matter content are depicted in Fig. \ref{combine fig3}, providing valuable insights into the properties of the system.

In Fig. \ref{fig6}, we observe the violation of the effective Null Energy Condition (NEC), indicating that in higher-order Gauss-Bonnet gravity, an alternative source $T_{\mu\nu}^{\mathrm{eff}}$ is present. This violation suggests the existence of exotic matter that enables the formation and maintenance of the microscopic wormhole.
Figure \ref{fig:isoropp} presents the behavior of the energy density $\rho$, while also illustrating the behavior of the quantity $\rho+P$, where $P$ represents the pressure. Notably, within the region $0.15 \leq r < 0.35$, both the NEC and the Weak Energy Condition (WEC) remain positive. This finding indicates the potential formation of a physically realistic and traversable microscopic wormhole supported by an isotropic fluid.

These results support the viability of the wormhole solution within the context of isotropic matter and higher-order Gauss-Bonnet gravity. The violation of the NEC, combined with the positive energy conditions in the relevant region, indicates the possibility of constructing a stable and traversable microscopic wormhole using the specific matter content under consideration.

\section{Equation of state for barotropic fluids}
In this section, we investigate wormhole geometries by employing a barotropic equation of state for both the radial and tangential pressures.
\subsection{Case of $P_{r}=k\rho$}
\begin{figure}[h]
\centering
\includegraphics[width=0.5\linewidth]{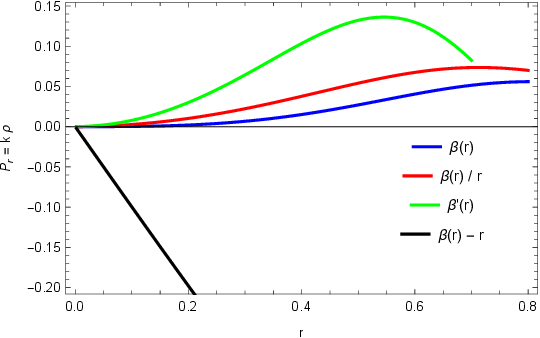}
\caption{Behavior of $\beta(r)$, $\beta(r)-r$, $\frac{\beta(r)}{r}$, and $\beta'(r)$ versus $r$ for $P_{r}=k\rho$.}
\label{combine fig4}
\end{figure}

\begin{figure}[h]
\centering

\subfloat[Behavior of effective energy conditions]{%
  \includegraphics[width=0.45\textwidth]{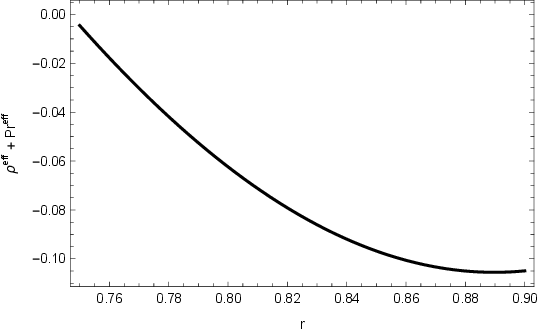}
  \label{fig:neceos}
}
\hspace{0.05\textwidth}
\subfloat[Behavior of energy conditions]{%
  \includegraphics[width=0.45\textwidth]{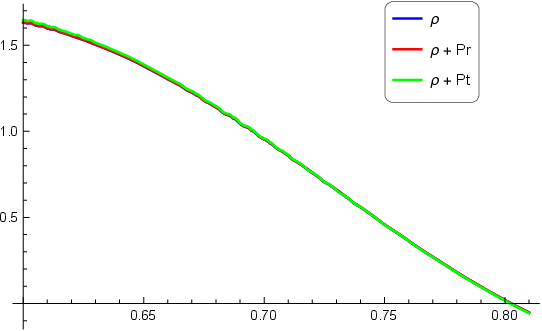}
  \label{fig:eosec}
}
\caption{Energy Condition Behavior for $P_{r}=k\rho$.}
\label{combine fig5}
\end{figure}

We consider $P_{r}=k\rho$, where $k$ represents the equation of state parameter. By utilizing this equation of state,
we find a nonlinear differential equations.
We numerically solve these equations to examine the behavior of $\beta(r)$, and the corresponding results are presented in Fig. \ref{combine fig4}. Which illustrates the positive and increasing evolution of $\beta(r)$. This shows the value of the wormhole throat radius, which is determined to be located at $r_0\approx0.02$ such that $\beta(r_0)=r_0$ and demonstrates that the behavior of $\beta(r)$ exhibits asymptotic flatness, and confirms that the constraint $\beta'(r)<1$ is satisfied at $r=r_0$.
Furthermore, Fig. \ref{combine fig5} illustrate the energy conditions in this scenario. In Fig. \ref{fig:neceos}, the behavior reveals the violation of the NEC for $T_{\mu\nu}^{\mathrm{eff}}$. On the other hand, Fig. \ref{fig:eosec} demonstrates the satisfaction of the NEC, as well as the WEC, for the ordinary matter distribution described by $\rho\geq0$, $\rho+P_{r}\geq0$, and $\rho+P_{t}\geq0$. Therefore, the wormhole geometries in this scenario are supported by ordinary matter.

\subsection{Case of $P_{t}=k\rho$}
\begin{figure}[h]
\centering
\includegraphics[width=0.5\linewidth]{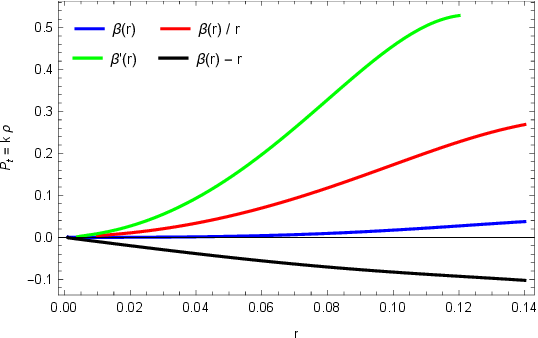}
\caption{Behavior of $\beta(r)$, $\beta(r)-r$, $\frac{\beta(r)}{r}$, and $\beta'(r)$ versus $r$ for $P_{t}=k\rho$.}
\label{combine fig6}
\end{figure}

\begin{figure}[h]
  \centering

  \subfloat[Behavior of effective energy conditions]{%
    \includegraphics[width=0.45\textwidth]{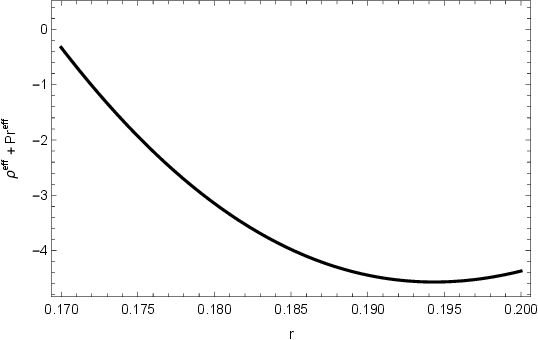}
    \label{fig:neceos2}
  }\hfill
  \subfloat[Behavior of energy conditions]{%
    \includegraphics[width=0.45\textwidth]{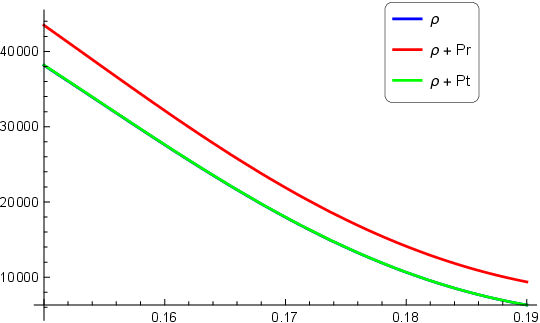}
    \label{fig:eosec2}
  }

  \caption{Energy Conditions for $P_{t}=k\rho$.}
  \label{combine fig7}
\end{figure}

When considering the tangential pressure, we opt for a barotropic equation of state in the format of $P_{t}=k\rho$. By analyzing the field equations, we arrive at a third-order differential equation with respect to $\beta(r)$, which allows us to investigate the potential presence of wormhole solutions.
This differential equation is tackled numerically, and the corresponding outcomes are depicted in Figure \ref{combine fig6}. The behavior of $\beta(r)$, $\frac{\beta(r)}{r}$, and $\beta'(r)$ follows a similar pattern to the previous scenarios. The throat of the wormhole is positioned around $r_0\approx0.01$, at which point the $\beta(r)-r$ curve converges toward zero.
We solve this equation numerically and present the corresponding results in Fig. \ref{combine fig7}. In which, Fig. \ref{fig:neceos2} illustrates the violation of the NEC in higher-order Gauss-Bonnet gravity throughout the evolution. On the other hand, the behavior of $\rho\geq0$, $\rho+P_{r}\geq0$, and $\rho+P_{t}\geq0$ demonstrate positive values, as shown in Figure \ref{fig:eosec2}. Consequently, a physically viable microscopic wormhole can be formed within this region.

\section{Stability Analysis}
In addition to ensuring traversability, it is imperative for the solutions to remain stable. This stipulation pertains to the stability of fluids within the throat while also adhering to the energy conditions outlined above. Therefore, we investigate the perturbation condition through the lens of the adiabatic sound speed, denoted as $V_s$. This notion draws an analogy to the definition of sound speed in the context of adiabatic perturbations, akin to its role in fluid dynamics. With this in mind, we establish the adiabatic sound speed as follows:
\begin{equation}
{V_s} = \sqrt {{{\left( {\frac{{\partial P}}{{\partial \rho }}} \right)}_s}}
\end{equation}
In our scenario, we encounter three distinct solution types: the anisotropic case, the isotropic case, and the barotropic case, each with its own set of sub-cases. To ascertain stability, we employ a sound speed analysis, where the range of sound speeds should conform to a specified boundary, typically denoted as [0,1].

The analysis results are visually presented in Fig. \ref{figs} below:

\begin{figure}[h]
  \centering
  \subfloat[Stability analysis of the anisotropic case.]{\includegraphics[width=0.45\linewidth]{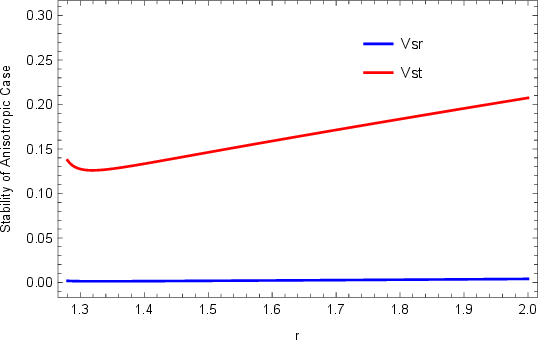}\label{s1}}
  \hfill
  \subfloat[Stability analysis of the isotropic case.]{\includegraphics[width=0.45\linewidth]{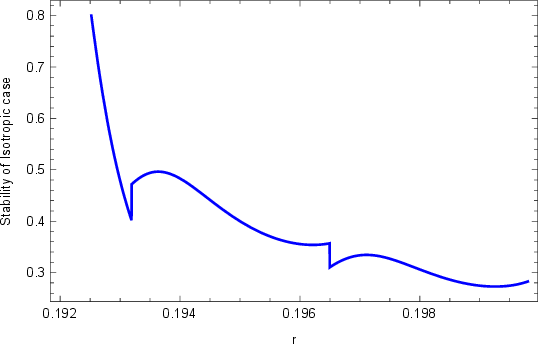}\label{s2}}

  \vspace{0.5cm}

  \subfloat[Stability analysis of the barotropic case with $P_r = k \rho$.]{\includegraphics[width=0.45\linewidth]{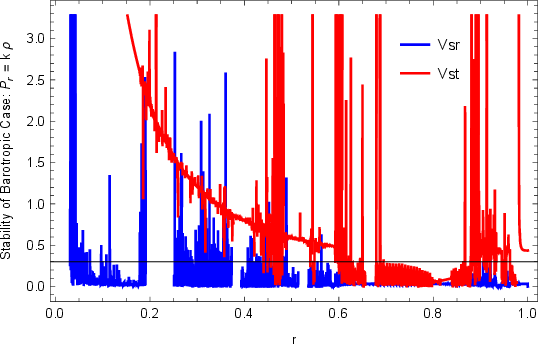}\label{s3}}
  \hfill
  \subfloat[Stability analysis of the barotropic case with $P_t = k \rho$.]{\includegraphics[width=0.45\linewidth]{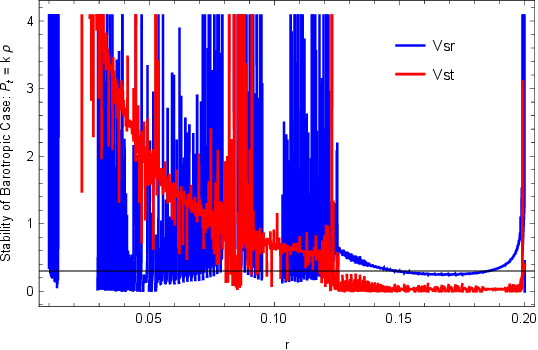}\label{s4}}

  \caption{Stability Analysis}
  \label{figs}
\end{figure}

In the Fig. \ref{figs}, we have segmented the analysis corresponding to a particular case:\\

\begin{itemize}
\item[1)] \textbf{Anisotropic Case (Figure \ref{s1}):} This subplot depicts the stability analysis of the anisotropic case. It can be observed that our solutions exhibit stability within this context.

\item[2)] \textbf{Isotropic Case (Figure \ref{s2}):} The next subplot showcases the stability analysis results for the isotropic case. Here too, our solutions demonstrate stability.

\item[3a)] \textbf{Barotropic Case: $P_r = k \rho$ (Figure \ref{s3}):} The stability analysis of the barotropic case with the relationship $P_r = k \rho$ is displayed. The results unveil regions of stability within this configuration.

\item[3b)] \textbf{Barotropic Case: $P_t = k \rho$ (Figure \ref{s4}):} Lastly, the stability analysis for the barotropic case, where $P_t = k \rho$, is exhibited. Similar to the previous barotropic scenario, this case also reveals certain regions where stability is maintained.
\end{itemize}

The analyses presented in Fig. (\ref{figs}) offer a compelling exploration of the underlying physics governing wormholes and compact objects within the realm of higher-curvature gravity. These analyses not only provide insights into their stability and behavior but also offer glimpses into how these structures respond to the intricate interplay between curvature modifications and their intrinsic properties. By unraveling these dynamics, we gain a deeper grasp of the fundamental physics at play in these exotic cosmic formations within the context of higher-curvature gravity.
In summary, the figures conclusively depict the stability of our solutions across different cases. The anisotropic and isotropic cases exhibit consistent stability throughout. For the barotropic cases, regions of stability are identifiable, thereby reinforcing the overall robustness of our solutions within varying conditions.
\section{Comparative Analysis: Wormhole Formation in Higher-Order Gauss-Bonnet Gravity vs. GR}
In this comparative analysis, we delve into the intriguing possibility of creating wormholes within two distinct gravitational frameworks: Higher-Order Gauss-Bonnet Gravity characterized by the Gauss-Bonnet invariant and the well-established G.R. Our exploration encompasses three distinct scenarios, each offering its unique set of characteristics and implications.

\begin{itemize}
  \item \textbf{Anisotropic Scenario}

Within the realm of anisotropy, a striking disparity emerges between the outcomes of Higher-Order Gauss-Bonnet Gravity and GR. In the former, the presence of the Gauss-Bonnet invariant induces profound alterations to the gravitational field equations, ushering in the tantalizing possibility of stable wormhole formation without the requirement for exotic matter. This transformation stems from the modification of the spacetime geometry, allowing anisotropic conditions to be satisfied using conventional matter content. Conversely, GR steadfastly demands the presence of exotic matter characterized by negative energy density to maintain the stability of wormhole structures-an imperative that often leads to scenarios marred by physically unrealistic properties.

  \item \textbf{Isotropic Scenario}

  Transitioning into the isotropic scenario, Higher-Order Gauss-Bonnet Gravity, equipped with the Gauss-Bonnet invariant, continues to exhibit its remarkable versatility by enabling the creation of stable wormholes using ordinary matter content. The spherical symmetry inherent in the geometry, coupled with the Gauss-Bonnet term, provides the essential prerequisites for wormhole formation without the necessity of exotic matter. In stark contrast, G.R, much like the anisotropic case, mandates the inclusion of exotic matter as it fails to naturally accommodate wormhole solutions with regular matter content, as dictated by the standard Einstein-Hilbert equations.

  \item \textbf{Equation of State Scenario}

  Our analysis extends further to scrutinize the implications of varying matter equations of state. Within this scenario, Higher-Order Gauss-Bonnet Gravity, endowed with the Gauss-Bonnet invariant, once again demonstrates its adaptability. It permits stable wormhole creation across a spectrum of matter equations of state, all of which feature regular matter components. This remarkable flexibility underscores Higher-Order Gauss-Bonnet Gravity's capacity to accommodate diverse energy conditions for wormhole creation. In sharp contrast, G.R maintains its unwavering demand for exotic matter, regardless of the equation of state under consideration. The requirement for exotic matter, characterized by its unconventional energy properties, remains a constant prerequisite for wormhole stability within General Relativity.
\end{itemize}
\begin{table}[h]
\centering
\caption{Wormhole Formation}
\label{tab:wormholes}
\begin{tabular}{|c|c|c|}
\hline
\textbf{}           & \textbf{Higher-Order Gauss-Bonnet Gravity}, $T_{\mu\nu}^{eff} = T_{\mu\nu}  + {T_{\mu\nu}^{GB}}$        & \textbf{GR (General Relativity)}, $T_{\mu\nu}^{eff} = T_{\mu\nu}$
\\ \hline
\textbf{Anisotropic} & Possible with Ordinary Matter & Only Possible with Exotic Matter   \\ \hline
\textbf{Isotropic}   & Possible with Ordinary Matter & Only Possible with Exotic Matter   \\ \hline
\textbf{Equation of State} & Possible with Ordinary Matter & Only Possible with Exotic Matter   \\ \hline
\end{tabular}
\end{table}
In summary, our comparative analysis unveils the distinctive attributes of Higher-Order Gauss-Bonnet Gravity, incorporating the Gauss-Bonnet invariant, which enables the potential creation of wormholes employing ordinary matter content across a multitude of scenarios. Conversely, GR consistently necessitates the presence of exotic matter endowed with unconventional energy properties to sustain wormhole stability. This juxtaposition underscores the potential significance of alternative gravitational theories in the realm of cosmological studies and their profound implications for fundamental physics.

%\newpage
\section{Conclusions}
Wormholes have been the subject of extensive investigation in various modified gravity theories, shedding light on their existence and characteristics. Prior research has explored wormhole geometries within $f(R)$ modified gravity \cite{Lobo:2009ip}, presented exact traversable wormhole solutions in bumblebee gravity \cite{Ovgun:2018xys}, and scrutinized wormhole solutions within symmetric teleparallel gravity \cite{Mustafa:2021ykn}. Innovations such as Einstein-scalar-Gauss-Bonnet wormholes without exotic matter \cite{Antoniou:2019awm}, wormhole formation in $f(R,T)$ gravity with varying Chaplygin gas and barotropic fluid \cite{Elizalde:2018frj}, and novel cosmological acceleration through a holographic wormhole \cite{Antonini:2022ptt} have broadened our understanding. Transitions from black holes to wormholes within scalar-tensor Horndeski theory \cite{Chatzifotis:2021hpg}, wormhole geometries in $f(Q)$ gravity \cite{Banerjee:2021mqk}, and wormhole solutions examined through the Poynting-Robertson effect in extended gravity \cite{DeFalco:2021klh} have further advanced our comprehension. The stationary generalizations for the Bronnikov-Ellis wormhole and the vacuum ring wormhole \cite{Volkov:2021blw}, as well as traversable wormholes within Einstein 3-form theory \cite{Bouhmadi-Lopez:2021zwt}, have all added to the mosaic of wormhole research, significantly contributing to our understanding of these enigmatic structures.
In this paper, we delve into the intriguing realm of higher-order Gauss-Bonnet gravity theory, exploring its profound implications for the fabric of spacetime. We focus on the field equations governing this theory, offering a novel perspective that extends beyond the conventional framework. Moreover, we introduce a compelling and viable model that encapsulates this theory's essence while delving into the intricate geometry of wormholes.
Within the scope of our study, we discern three distinct yet interconnected scenarios, each offering unique insights into the behavior of spacetime curvature. In the first scenario, we delve into the anisotropic realm, ingeniously crafting a specific shape function that adheres rigorously to the foundational conditions required for wormhole construction. Remarkably, we uncover regions within this anisotropic landscape where traversable wormholes can indeed materialize, propelled by ordinary matter content.
Our exploration advances further as we transition into the second scenario, characterized by a perfect fluid assumption, namely, assuming radial and tangential pressures to be equivalent ($P_r = P_t$). We embolden this scenario with the formidable machinery of higher-order Gauss-Bonnet gravity, meticulously solving the corresponding field equations. In doing so, we uncover a profound connection between the shape function, denoted as $\beta(r)$, and the feasibility of constructing a wormhole. This revelation cements the intricate interplay between the curvature of spacetime and the structural integrity of wormholes.
The third scenario unfurls with the introduction of a barotropic equation of state, imparting an added layer of complexity to our investigation. Employing numerical techniques, we unravel the differential equations that govern this scenario, thereby unearthing the shape function that governs the construction of a wormhole. A rigorous examination of this shape function ensues, scrutinizing its compliance with the stringent conditions required for the manifestation of wormholes.
Intriguingly, our inquiry extends beyond the realm of mere construction, delving into the stability of the solutions we unearth. Each of the three scenarios is subjected to meticulous stability analysis, unearthing the delicate equilibrium between gravitational forces and matter content. This comprehensive investigation stands as a testament to the robustness of our findings, fortifying the credibility of the potential wormhole geometries we reveal.
Summarily, our manuscript culminates in a resonant conclusion that resonates with the symphony of modified gravity and wormhole physics. We have embarked on a journey to explore static spherically symmetric wormholes within the tapestry of higher-order Gauss-Bonnet gravity. By meticulously crafting the intricate interplay between curvature and matter, we have derived a model that bridges the gap between theory and physical reality. Notably, we have discovered that the violation of the Null Energy Condition (NEC) is facilitated by the effective energy-momentum tensor $T^\text{(eff)}_{\alpha\beta}$, echoing similar phenomena observed in other modified gravity theories. The interplay between the NEC and the Weak Energy Condition (WEC) for $T^\text{(m)}_{\alpha\beta}$ has also been rigorously examined, revealing intriguing insights into the fabric of spacetime.
A significant facet of our study is the construction of wormhole solutions through numerical computations. Our findings, remarkably, showcase the delicate balance between theoretical constructs and physical viability. Particularly noteworthy is the revelation that while NEC and WEC violations are intrinsic to traceless, isotropic, and tangential barotropic fluids, a fascinating exception arises for the radial barotropic fluid in specific spatial regions. This newfound equilibrium, where energy conditions harmonize with wormhole geometries, underscores the profound influence of higher curvature terms in sustaining these enigmatic passages in spacetime.
In past work, wormholes in various modified gravity theories were investigated, each contributing valuable insights into their existence and characteristics. These studies explored wormhole geometries in specific gravitational frameworks and presented exact solutions for traversable wormholes. While they made significant progress in understanding these intriguing cosmic structures, there were some limitations to their findings.
In contrast, our study embarks on a pioneering journey into the realm of traversable wormholes within the framework of the higher-curvature gravity model, as depicted in Eq. (\ref{model}). What truly distinguishes our work from previous studies is the meticulous examination of the energy-momentum tensor, consistently revealing violations of the null energy condition across the wormhole throat for all scenarios. This discovery is a key highlight, as it suggests the possibility of constructing stable wormholes with conventional matter content, a feat previously thought to be restricted to specific cases, such as the radial barotropic case.
Furthermore, we integrate a diverse range of matter content, encompassing anisotropic, isotropic, and barotropic fluids. This broadens the practical applicability of our results and showcases the versatility of our approach. We also unveil a novel pattern of expansion and contraction within these regions, shedding light on previously unexplored facets of traversable wormholes in the context of modified gravity.
In conclusion, our comparative analysis between Higher-Order Gauss-Bonnet Gravity and General Relativity has illuminated stark differences in their capabilities regarding wormhole formation. Higher-Order Gauss-Bonnet Gravity has demonstrated remarkable adaptability, allowing for the creation of stable wormholes with ordinary matter content across various scenarios. In contrast, General Relativity consistently demands the presence of exotic matter with unconventional energy properties to uphold wormhole stability. These findings extend beyond theoretical curiosity and hint at the potential significance of alternative gravitational theories in the realm of cosmological studies and their profound implications for fundamental physics. In essence, our research represents a significant leap forward in understanding the possibilities and limitations inherent in these two gravitational frameworks.
In essence, our work unifies the tapestry of higher-order Gauss-Bonnet gravity, wormhole physics, and the intricate dance of energy conditions. Through meticulous analysis and groundbreaking insights, we have not only illuminated the potential for physically plausible wormholes but also affirmed the pivotal role that higher curvature terms play in upholding their existence. As the final note in this symphony of exploration, our conclusions stand as a testament to the ceaseless wonders that emerge at the confluence of theoretical innovation and empirical discovery.

%%%%%%%%%%%%%%%%%%%%%%%%%%
\section*{Acknowledgments}

The work of KB was partially supported by the JSPS KAKENHI Grant
Number 21K03547.

%%%%%%%%%%%%%%%%%%%%%%%%%%

\end{document}